\documentclass[12pt]{iopart}
\usepackage{graphicx}

\usepackage{layouts}
\usepackage[utf8]{inputenc}
\usepackage{graphicx}
\usepackage{cite}
\usepackage{float}
\expandafter\let\csname equation*\endcsname\relax
\expandafter\let\csname endequation*\endcsname\relax
\usepackage{amsmath,amssymb}
\usepackage{iopams}
\usepackage{IEEEtrantools}
\usepackage{hyperref}
\usepackage{xcolor}

\newcommand{\jdes}{J_{\text{des}}}
\newcommand{\jexp}{J_\text{exp}}

\newcommand{\jdesTilde}{\tilde{J}_{\text{des}}}
\newcommand{\jexpTilde}{\tilde{J}_\text{exp}}

\newcommand{\jk}{J^{(k)}}

\newcommand{\ket}[1]{|#1\rangle}

\newcommand{\sx}{\sigma_x}

\newcommand{\down}{\ket{\!\!\downarrow}}
\newcommand{\up}{\ket{\!\!\uparrow}}
\newcommand{\I}{\mathcal{I}}
\newcommand{\com}{center-of-mass~}
\DeclareMathOperator{\diag}{diag}

\graphicspath{ {./Figures/} }

\begin{document}

\title{Interaction graph engineering in trapped-ion quantum simulators with global drives}
\author{Antonis Kyprianidis$^1$, A. J. Rasmusson$^1$ , Philip Richerme$^{1,2}$}
\address{$^1$Indiana University Department of Physics, Bloomington, Indiana 47405, USA}
\address{$^2$Indiana University Quantum Science and Engineering Center, Bloomington, Indiana 47405, USA}

\begin{abstract}
Trapped-ion quantum simulators have demonstrated a long history of studying the physics of interacting spin-lattice systems using globally addressed entangling operations. Here, we seek to broaden and delimit the classes of effective spin-spin interactions achievable using exclusively global driving fields. We find that new categories of interaction graphs become achievable with perfect or near-perfect theoretical fidelity by tailoring the coupling to each vibrational mode of the ion crystal, or by shaping the trapping potential to include specific anharmonic terms. We also derive a rigorous test to determine whether a desired interaction graph is accessible using only globally driven fields. These tools broaden the reach of trapped-ion quantum simulators so that they may more easily address open questions in materials science and quantum chemistry.

\end{abstract}
\maketitle

\section{Introduction}

For over 15 years trapped-ion platforms have pushed forward the frontier of quantum simulation, wherein a controlled quantum system is made to emulate the behavior of a target system \cite{feynman1982,monroe2021programmable}. Trapped ions exhibit key features for quantum simulation, such as the ability to form lattices with long quantum coherence times \cite{wang2021single}, near-perfect state preparation and measurement \cite{olmschenk2007manipulation,noek2013high}, and high-fidelity quantum operations \cite{ballance2016high} that can be controlled and reprogrammed using laser light. As a result, trapped-ion systems have been used to simulate diverse problems from condensed-matter physics \cite{monroe2021programmable} and quantum chemistry \cite{hempel2018quantum,gorman2018engineering,nam2020ground,whitlow2022simulating,richerme2023quantum,valahu2023direct} to high-energy physics \cite{martinez2016real} and cosmology \cite{schuster2022many}.

In principle, trapped ions can simulate any possible quantum system since standard single-qubit rotations \cite{ballance2016high,gaebler2016high} and two-qubit entangling operations \cite{molmer1999multiparticle} form a universal gate set. However, there is no guarantee that a target system of interest can be \emph{efficiently} represented by a sequence of quantum gates. For instance, simulating the quantum dynamics of a generic Hamiltonian $H$ requires decomposing its unitary propagator $U=e^{-iHt/\hbar}$ into an exponential number of 2-qubit gates ($\mathcal{O}(4^N)$ for an $N$-qubit system \cite{shende2004minimal}). To address this intractability, algorithms for Hamiltonian simulation have been developed which exploit inherent symmetries and properties of certain types of Hamiltonians to approximate their evolutions using far fewer quantum gates \cite{berry2015simulating,low2017optimal}. Alternatively, non-gate-model techniques of analog and digital quantum simulation \cite{georgescu2014quantum} have long-demonstrated success by simulating Hamiltonians which inherit the native interactions of the underlying trapped-ion system \cite{monroe2021programmable,lanyon2011universal,kyprianidis2021observation}.

To date, analog and digital Hamiltonian simulation methods have largely centered around global Ising-type $X\!X$-interactions, which arise natively through the application of a M{\o}lmer-S{\o}rensen operation \cite{molmer1999multiparticle,porras2004effective}. Such interactions are most commonly used to generate long-range Ising couplings between effective quantum spins which decay algebraically with distance \cite{kim2009entanglement}. This simple foundation, when combined with the ability to apply effective magnetic fields, has led to over a decade of trapped-ion experiments probing classical and quantum Ising models, $XY$ and Heisenberg models, open quantum systems, and non-equilibrium physics \cite{monroe2021programmable}.

In this work, we seek to broaden and delimit the classes of spin-spin couplings achievable using globally addressed M{\o}lmer-S{\o}rensen interactions. We find that a wide range of native interactions becomes accessible by controlling the participation of each vibrational mode of the lattice during a simulation, or by purposefully confining ions within anharmonic trapping potentials; neither approach requires locally addressed entangling gates. These techniques would enable straightforward experiments probing, for instance, spin models with pure nearest-neighbor interactions, ring topologies, infinite-range couplings, higher-dimensionality spin lattices, and multipartite quantum systems with interacting degrees of freedom.

The article is structured as follows. Section \ref{framework} reviews the standard framework of generating effective spin-spin Ising interactions in trapped ions using global laser beams. In Sec. \ref{multimode}, we show how this standard treatment may be extended by adding multiple frequency components to the global beams, unlocking new classes of native interaction profiles. Section \ref{TrappingPotential} presents further classes of interactions which may be generated by modifying the trapping potential experienced by the ions. We conclude in Sec. \ref{challenges} with a discussion of potential error sources, challenges, and opportunities for experimental implementation.

%%%%%%%%%%%%%%

\section{Framework for Effective Spin-Spin Interactions Within Ion Coulomb Crystals}
\label{framework}

%%%%%%%%%%%%%

\subsection{The Motional Mode Structure of Ion Coulomb Crystals}

A collection of $N$ atomic ions, when confined in a Paul trap using electric fields and cooled to milliKelvin temperatures, forms a Coulomb crystal with $3N$ vibrational modes of motion \cite{NistBible}. Ion Coulomb crystals may be created in one, two, or three dimensions depending on the number of ions and the configuration of trapping voltages \cite{NistBible, kaufmann2012precise, d2021radial,qiao2021double,kiesenhofer2023controlling,hornekaer2002formation,thompson2015ion}. In this section, we first consider the motional modes for a one-dimensional ion chain, in which the confinement along the chain axis $z$ is weak compared to the confinement along the two transverse directions. Later, we generalize to Coulomb crystals in higher dimensions.

If the transverse confinement along the $x$ and $y$ axes is harmonic, and the axial potential is symmetric around the trap center, the potential energy of a single ion is
\begin{equation}
\label{eq:potential}
U_\text{trap}(x,y,z) = \frac{1}{2}m \left(\omega_{x}^2 x^2 + \omega_{y}^2 y^2 + \tilde{\omega}_{z}^2 \sum_{n=2}^\infty \beta_n z^n \right)
\end{equation}
where $m$ is the ion mass and $\omega_{x}, \omega_{y}$ are the motional \com (COM) frequencies along the $x$ and $y$ directions. In the simplest and most common case, the axial potential is also harmonic: $\beta_n=\delta_{n,2}$ and $\tilde{\omega}_{z}$ is equal to the COM mode frequency $\omega_z$ along the $z$ direction. For anharmonic axial potentials, the terms $\beta_n$ provide the contribution of each polynomial order and $\tilde{\omega}_{z}$ sets the overall numerical scale.

To compute the normal mode frequencies and amplitudes along the $x$ direction, we include the inter-ion Coulomb interaction terms and expand the potential in the $x$ direction to second order about the ions' equilibrium positions, with displacements $\xi_i$, $i=1,\ldots,N$  \cite{james1998quantum,Marquet2003}:
\begin{equation}\label{eq:Lx}
U^{(x)} \approx \frac{1}{2}m\tilde{\omega}_{z}^2 \sum_{i,j=1}^N A_{ij}\xi_i\xi_j.
\end{equation}
where the matrix $A_{ij}$ for transverse motion along the $x$ direction is \cite{Marquet2003}
\begin{equation}\label{eq:Aij}A_{ij} = 
\begin{cases}
\displaystyle \left(\frac{\omega_{x}}{\tilde{\omega}_{z}}\right)^2 
-\sum_{\substack{p=1\\p\neq i}}^N \frac{1}{|\vec{u}_i-\vec{u}_p|^3}\quad& \text{if}~i=j\\[7mm]
\displaystyle \frac{1}{|\vec{u}_i-\vec{u}_j|^3} \quad&\text{if}~i\neq j.
\end{cases} 
\end{equation}
Each vector coordinate $\vec{u}_i$ is the equilibrium position of the $i$th ion, made unitless by the length scale $l\equiv [q^2/(4\pi\epsilon_0 m \tilde{\omega}_{z}^2)]^{1/3}$; $q$ is the ion charge; and $\epsilon_0$ is the vacuum permittivity.

\begin{figure*}[hbtp]
\centering
\includegraphics[width=\textwidth]{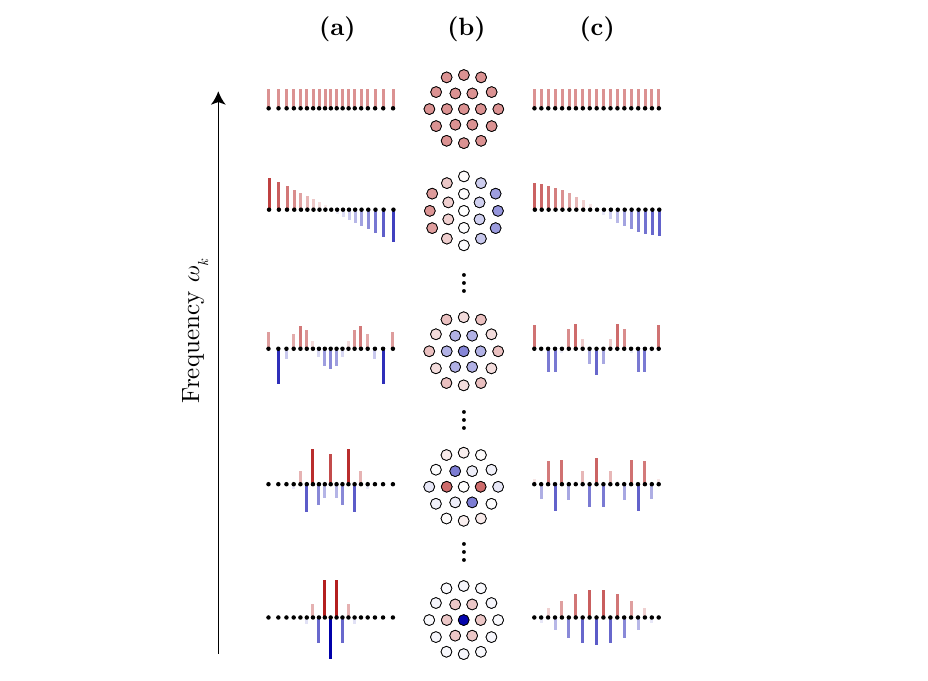}
\caption{
Sample transverse mode participation vectors for crystals of $N=19$ ions.
The first row shows the highest frequency mode, while the last row is the lowest frequency mode.
(a)  Harmonically confined 1D chain.
(b) Transverse mode participations for a 2D ion array.
The colored disks indicate the participation vectors for each ion using the same color scale as in (a).
(c) Mode participation vectors for an equispaced 1D chain. Compared to the harmonic case, the edge ions participate more strongly in the lower-frequency modes.
}
\label{fig:1}
\end{figure*}

The eigenvalues and eigenvectors of matrix $A_{ij}$ determine the set of $N$ transverse mode frequencies along the $x$ direction and participation vectors $\vec{b}_k$, $k=1,\ldots,N$.
The participation vector $\vec{b}_k$ contains the amplitude of oscillation of each ion $i$, $B_{ik}$, 
including its sign, with the normalization
\begin{equation}\label{eq:BNormalization}
\sum_i^N B_{ik}^2 = 1\ \forall k,\quad \sum_k^N B_{ik}^2 = 1\ \forall i.
\end{equation}
In the transverse direction, the highest-frequency mode (corresponding to the largest eigenvalue of $A_{ij}$) is the COM motion for which each ion participates with equal amplitude and sign. Lower-frequency eigenvectors contain an increasing number of nodes, with the lowest (zig-zag mode) exhibiting a sign flip for each adjacent ion (figure \ref{fig:1}(a)). Furthermore, the ion participation amplitudes for each mode feature an envelope which narrows more and more prominently at the chain center for lower-frequency modes, leaving the edge ions with nearly zero participation in the zig-zag motion. As we later show, the details of these mode participations inform the classes of spin-spin interactions accessible using global drives.
%This arises largely due to ions clustering at the center when confined in a harmonic axial potential, leading to preferentially larger Coulomb interactions near the center of the trap.

The above analysis may be extended to understand the transverse (drumhead) modes of a two-dimensional Coulomb crystal. This 2D ion geometry arises when the potential along two of the axes is weak compared to the third \cite{richerme2016two,d2021radial}; here, we again assume that confinement is harmonic in all three directions. For the 2D case, the $A_{ij}$ matrix is readily adjusted by incorporating the new ion equilibrium positions $\vec{u}_i$ in equation (\ref{eq:Aij}). The transverse modes of the 2D crystal (figure~\ref{fig:1}(b)) are found to share qualitative features as those of the 1D chain (figure~\ref{fig:1}(a)). For instance, the highest-frequency vibration is an equal-participation COM mode while the lowest-frequency vibration is a zig-zag mode strongly peaked near the center of the crystal, where adjacent concentric ``rings'' of ions oscillate out of phase.

When the axial trapping potential is made anharmonic, the corresponding transverse motional mode vectors may differ substantially from the harmonic case, even though the transverse axes remain harmonically confined. A notable example is an axial potential tailored such that the ions are equally spaced. In this case, the mode eigenvectors exhibit a more uniform distribution of motional amplitudes, particularly for the low-frequency modes (figure~\ref{fig:1}(c)). This configuration of equally spaced ions in anharmonic potentials will prove favorable for expanding the accessible quantum simulation experiments with global beams, as we will show in Section~\ref{TrappingPotential}.

%%%%%%%%%%%%%%%%%%%%%%%%%%%%%%%%%%%%%%%%%%%%%%

\subsection{Generating spin-spin couplings from laser--ion interactions}
\label{FromLaserIonInteractionsToSpinSpinInteractions}

Here we return to the 1D case to present a matrix formulation of laser-driven spin-spin couplings which will allow for simplified engineering of interaction graphs between trapped-ion qubits. We will consider the use of one set of $N$ transverse vibrational modes, with frequencies $\omega_k$, $k=1,\ldots,N$ to mediate effective spin-spin interactions within the crystal. We note that a parallel analysis may be performed for the axial modes which are likewise subject to the normalization conditions in equation (\ref{eq:BNormalization}).

Within each ion, a spin-1/2 qubit may be encoded in two electronic states $\down_z$ and $\up_z$ separated by $\hbar\omega_0$. Under the application of a bichromatic electric field of the form $\Vec{E}=E_0\hat{y}\cos[kx-(\omega_0\pm \mu) t+\phi]$ \cite{NistBible}, the Hamiltonian describing the laser-ion interaction for spin-1/2 systems may be written:
\begin{equation}\label{eq:MotherEqnMultiIon}
H_\text{phys} = \sum_i^N-d_i E_0 \sigma_x^i \cos (kx_i-\omega_0 t \pm \mu t+\phi)
\end{equation}
where $d_i$ is the magnitude of the electric dipole operator for the $i$th ion and $\mu$ is the detuning of the exciting radiation from the qubit transition $\omega_0$. The $\sx^i$ operator in equation (\ref{eq:MotherEqnMultiIon}), equivalent to the Pauli $X$ operator on ion $i$, arises from writing the dipole operator as a matrix coupling $\down_z$ and $\up_z$.

The time evolution of $H_\text{phys}$ may be approximated by evolution under an effective spin Hamiltonian,

\begin{equation}\label{eq:SpinHamiltonian}
H_\text{spin} = \sum_{i<j} J_{ij} \sx^i \sx^j
\end{equation}
where $J_{ij}\in \mathbb{R}$ is the effective spin-spin coupling between ions $i$ and $j$. This Ising-type Hamiltonian, with pairwise $\sx^i\sx^j$ interactions, emerges when the time evolution of $H_\text{phys}$ is written using the Magnus expansion \cite{monroe2021programmable,kim2009entanglement} in the regime where the motional modes are only virtually excited. Time evolution of $H_\text{spin}$ generates entanglement between coupled qubit pairs and is the foundation of nearly all quantum simulation experiments with trapped ions \cite{monroe2021programmable}.

When global driving fields (such as laser beams) are applied to the ions, the profile of the interaction matrix $J_{ij}$ is uniquely determined from the mode vectors $\vec{b}_k$. In equation (\ref{eq:SpinHamiltonian}), the coupling coefficients $J_{ij}$ are
\begin{equation}\label{eq:Jij}
J_{ij} = \Omega^2 R\sum_k^N \frac{B_{ik}B_{jk}}{\mu^2-\omega_k^2}
\end{equation}
where $\Omega$ is the global, on-resonance Rabi frequency at each ion, $R$ is the recoil frequency $R=\hbar(\Delta k)^2/(2m)$, and $\hbar\Delta k$ is the momentum transfer from the electric field to each ion. 

\begin{figure*}[tbp]
\centering
\includegraphics[width=\textwidth]{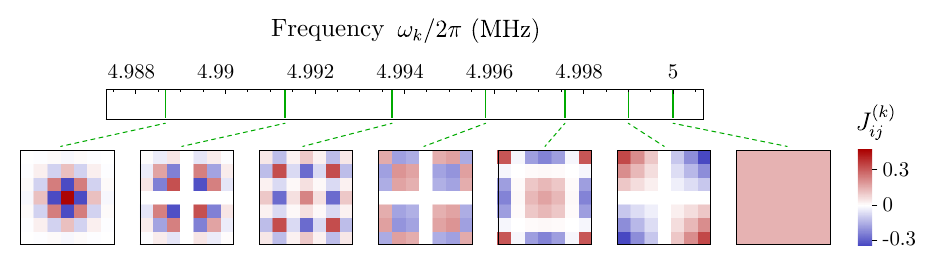}
\caption{Transverse mode spectrum of $N=7$ harmonically confined ions in a one-dimensional chain. Each mode frequency is associated with a $N \times N$ matrix $\jk$  which reflects the structure of normal mode vector $\vec{b}_k$.}
\label{fig:FromMotionalModesToInteractionGraphs}
\end{figure*}

To highlight the crucial role played by the mode eigenvectors in dictating the spin-spin interactions, we define the mode interaction matrices $\jk \equiv \vec{b}_k \otimes \vec{b}_k$, with matrix elements
\begin{equation}\label{eq:jkDefinition}
\jk_{ij}\equiv B_{ik} B_{jk}
\end{equation}
and rewrite Eq. (\ref{eq:Jij}) as
\begin{equation}\label{eq:LinearCombinationJks}
J = \sum_k^N c_k \jk.    
\end{equation}
Each $\jk$ matrix, which only depends on that mode's vector $\vec{b}_k$, is weighted by a coefficient \mbox{$c_k\equiv \Omega^2 R/(\mu^2-\omega_k^2)$} that depends on mode and laser frequencies and captures each mode's contribution to the final interaction matrix $J$. Sample $\jk$ matrices for a 1D chain of 7 ions are shown in figure \ref{fig:FromMotionalModesToInteractionGraphs}. Recasting equation (\ref{eq:Jij}) into the form of equation (\ref{eq:LinearCombinationJks}) highlights that, given a geometric configuration of trapped ions, the weights $c_k$ are the sole experimental knob available for interaction engineering with global beams. To gain full control over these weights, one may apply $M$ bichromatic tones, with frequency $\mu_m$ and amplitude $\Omega_m$, such that the sum over all contributions provides the desired $c_k$ \cite{Shapira2020}:
\begin{equation}
c_k = \sum_m^M c_k^{(m)}
\end{equation}
where $c_k^{(m)}\equiv \Omega_m^2 R/(\mu_m^2-\omega_k^2)$.

The normalization of the mode vectors (equation (\ref{eq:BNormalization})) leads to the property (proven in \ref{app:proofofjk}):
\begin{equation}\label{eq:JkProperty}
\sum_{k=1}^N \jk = \mathbf{1}.
\end{equation}
This implies that if all motional modes were equally driven by the exciting radiation, the resulting spin-spin interactions would be zero.
Likewise, this also implies that there are multiple sets $\{c_k\}$ which generate the same interaction model.
For instance, an equal-amplitude all-to-all coupling may be generated by coupling only to the COM mode, or by coupling to all modes \emph{except} the COM mode.
These properties of the $J^{(k)}$ matrices provide significant flexibility when experimentally implementing spin-spin couplings for quantum simulations with global beams.

%%%%%%%%%%%%%%%%%%%%%%%%%%%

\subsection{Metrics for Quantum Simulation Fidelity}
\label{sec:metrics}

For a generic physically inspired Hamiltonian of the form in equation (\ref{eq:SpinHamiltonian}), we may ask how closely a trapped-ion system with global beams replicates the desired spin-spin interactions.
Following \cite{Shapira2020}, we define the coupling matrix infidelity $\mathcal{I}$ to quantify the difference between a desired coupling matrix $\jdes$ and its best experimentally realizable approximation $\jexp$:
\begin{equation}
\label{eq:infidelity}
\mathcal{I} \equiv \frac{1}{2} \left( 1 - \frac{\langle \jexpTilde,\jdesTilde\rangle}{\| \jexpTilde\| \|\jdesTilde \|} \right)
\end{equation}
where we make use of the Frobenius matrix product and matrix norm, respectively
\[
\langle A,B\rangle = \Tr(A B) = \sum_{i,j}^N A_{ij} B_{ij}\quad\text{and}\quad\|A\| = \sqrt{\langle A,A \rangle}
\]
for $N\times N$ real symmetric matrices $A$, $B$.
$\tilde{A}$ denotes matrix $A$ with its diagonal subtracted.
This is necessary because the diagonal entries of the interaction matrices do not bear any physical significance and hence should not affect the infidelity measure.
The smallest value of $\mathcal{I}$ is 0 if $\jexp=r\jdes$, and the largest is 1 if $\jexp=-r\jdes$, with a scaling factor $r>0$.
For uncorrelated $\jexp$ and $\jdes$ , $\mathcal{I}=0.5$ on average.
In practice, if $\mathcal{I} \gtrsim 0.05$, the experimentally achievable $\jexp$ provides only a poor approximation of the desired $\jdes$.

\begin{figure}[tp]
    \centering\includegraphics[width=\textwidth]{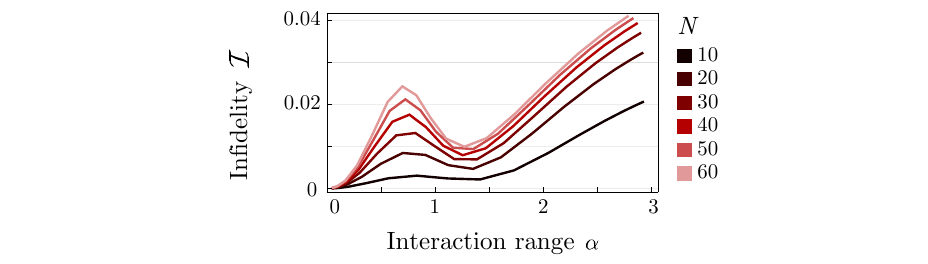}
    \caption{Infidelity of implementing the power law model of equation (\ref{eq:powerlaw}) for chains of $N=10$ to $60$ ions using a single bichromatic tone.
    Uniform all-to-all interactions ($\alpha$ = 0) are possible with perfect fidelity only in the limit of infinitely small detuning from the COM mode.
    For detunings where the tilt mode contributes strongly to the overall coupling matrix, as is the case for $0.5 \lesssim \alpha \lesssim 1$, the interaction profile deviates from power-law behavior by several percent.
    }
    \label{fig:3}
\end{figure}

As an example, we calculate the infidelity for one of the most commonly studied interaction graphs in ion-trap simulators: anti-ferromagnetic (AFM) Ising interactions that decay algebraically with distance \cite{monroe2021programmable}.
Experimentally, this model is typically realized by driving a bichromatic laser tone $\mu > \omega_\text{COM}$ which couples most strongly to the transverse COM mode and most weakly to the zig-zag mode.
The resulting spin--spin interactions resemble a power law
\begin{equation}
\label{eq:powerlaw}
\jexp = \Omega^2 R\sum_k^N \frac{B_{ik}B_{jk}}{\mu^2-\omega_k^2}\approx \frac{J_0}{|i-j|^\alpha}
\end{equation}
where $J_0>0$ and the interaction range $\alpha$ is experimentally tunable between 0 and 3 \cite{britton2012engineered}. For small numbers of ions, and for limited interaction ranges $0.5 < \alpha < 2$, the approximation in equation (\ref{eq:powerlaw}) has been sufficient to study a variety of interesting physical phenomena \cite{monroe2021programmable}, such as the ground-state \cite{islam2013emergence} and dynamical \cite{richerme2014non} properties of power-law AFM spin lattices. However, as shown in figure~\ref{fig:3}, deviations from the power-law model are significantly larger when considering the full range of possible $\alpha$ and larger numbers of ions. Accessing these regimes with high fidelity therefore requires the development of new experimental techniques to more precisely engineer the desired interaction graph.

%%%%%%%%%%%%%%%%%%%%%%%%%%
%%%%%%%%%%%%%%%%%%%%%%%%%%

\section{Expanded set of interaction profiles using multiple modes}
\label{multimode}

Following the framework introduced in Section \ref{framework}, here we delineate the full range of interaction profiles which may be accessed using global beams applied to harmonically confined ions.
We present a rigorous test to determine whether a desired set of couplings $\jdes$ may be experimentally realized, and we discuss several example spin models which may be simulated without requiring locally-addressed entangling gates.
Finally we show how seemingly inaccessible coupling matrices $\jdes$ may be recast, without affecting their underlying physical properties, to improve their implementation fidelity.

%%%%%%%%%%%%%%%%%%%%%%%%%%
\subsection{Accessible interaction graphs}
%%%%%%%%%%%%%%%%%%%%%%%%%%
%Scalable, high-fidelity simulations of such coupled quantum systems will therefore require new experimental approaches. To that end, the interaction engineering techniques described below.

%New experimental techniques will therefore be required to achieve scalable, high-fidelity simulations of such coupled quantum systems. In the following sections, a new suite 

We begin with the observation that a long-range AFM Ising model with pure power-law decays, introduced in Section \ref{sec:metrics}, cannot be realized perfectly using only global beams.
Rather, its experimental implementation is approximate, relying on the specific mode couplings which arise from, for example, a single bichromatic tone of frequency $\mu > \omega_{\text{COM}}$. These couplings shape the weights $c_k$ with which the $\jk$ matrices are summed together, serendipitously resulting in a $\jexp$ which resembles an Ising model with power law interactions.

To systematically determine which interaction profiles are accessible with theoretically perfect fidelity using global beams, we return to equation (\ref{eq:LinearCombinationJks}). Given a collection of ions in a harmonic potential, the mode eigenvectors $\vec{b}_k$ are uniquely determined, which in turn determines the $\jk$ matrices. Hence the only free parameters in equation (\ref{eq:LinearCombinationJks}) are the mode weights $c_k$; controlling these then opens the pathway to engineering a desired coupling matrix $\jdes$. Already, experiments have demonstrated that a desired set of weights $\{c_k\}$ may be applied using $M$ multiple bichromatic tones, each with independent frequencies $\mu_m$ and amplitudes $\Omega_m$ \cite{shapira2018robust,Shapira2020}. However, we note that only a subset of arbitrary interaction matrices $J$ can be realized if the experimental apparatus uses global laser beams. This is because there are of order $\mathcal{O}(N^2)$ free parameters in an arbitrary $J$ matrix, whereas the mode interaction matrices $\jk$ provide only $\mathcal{O}(N)$ linearly independent degrees of freedom.

Our key result in this section, proven in \ref{app:AccessibilityProof}, applies to any set of vibrational modes along a harmonically confined axis used for quantum simulation.
Under this condition, a desired interaction matrix $\jdes$ is accessible with theoretically perfect fidelity if and only if it is diagonalized by the mode vector matrix $B$:
\begin{equation}\label{eq:Centerpiece}
\jdes\text{ is accessible}\ \iff\  C\equiv B^T\jdes B\text{ is diagonal}
\end{equation}
Equation (\ref{eq:Centerpiece}) holds so long as the diagonal entries of $\jdes$, otherwise bearing no physical significance, are chosen to zero the sum of each row and column (i.e. $\jdes$ is written in graph Laplacian form \cite{West2001}).
If $\jdes$ is accessible, the diagonal matrix $C$ contains the weights $c_k$ of each mode interaction matrix $\jk$ necessary to realize $\jdes$, up to an additive constant (arising from equation (\ref{eq:JkProperty})). Thus, for accessible $\jdes$ matrices,
\begin{equation}
\label{eq:jdesck}
\jdes = \sum_{k=1}^N c_k \jk.
\end{equation}
Equation (\ref{eq:Centerpiece}) formalizes the intuition that only interaction matrices compatible with the structure of the crystal's motional modes can be realized.
We note that not all spatial axes need harmonic confinement for equation (\ref{eq:Centerpiece}) to hold; it is valid as long as the axis used for the entangling operations is harmonically confined.

Another property of realizable interaction matrices, specific to 1D ion chains with symmetric confining potentials, is their symmetry with respect to their anti-diagonal.
That is, two pairs of qubits mapped to each other by a reflection around the chain center have the same interaction strength:
\begin{equation}
\jk_{ij} = \jk_{N-i+1,N-j+1}\quad  i,j=1,2,\ldots,N.
\end{equation}
This follows from equation (\ref{eq:jkDefinition}) and the fact that in a symmetric potential, all resulting $\jk$ matrices are reflection-symmetric about their anti-diagonal (see figure \ref{fig:FromMotionalModesToInteractionGraphs}).
Therefore, the part of a desired interaction matrix $\jdes$ that is not symmetric with respect to the anti-diagonal cannot be simulated.

A final important property is that any linear combination of accessible interaction graphs is also accessible. This arises as a consequence of the linearity of equation (\ref{eq:jdesck}) and is applicable to ion crystals in any dimension. As we will show in Section \ref{ExactlyRealizableGraphs}, this may be leveraged to build increasingly complex interaction profiles from combinations of simple interaction graphs.

%%%%%%%%%%%%%%%%%%%%%
\subsection{Example Applications: Exactly Realizable Interaction Graphs}
\label{ExactlyRealizableGraphs}
%%%%%%%%%%%%%%%%%%%%%%
\subsubsection{All-to-All Interactions}
As a first example, we consider $N$ spins interacting with an equal-magnitude, all-to-all interaction. This Hamiltonian allows for analog simulation of the Ising model (equation (\ref{eq:powerlaw})) with an interaction range $\alpha=0$. In addition, all-to-all interactions provide a pathway for speeding up certain algorithms in the quantum gate model, since they can more efficiently implement operations such as Toffoli gates, Quantum Fourier Transforms, or GHZ state creation \cite{ivanov2015efficient,martinez2016compiling,maslov2018use,lu2019global}.

Realization of this model is not feasible using the common approach of applying a single bichromatic tone. In principle, one might tune that tone very close to the COM mode such that it dominates over the contributions from all other modes. In practice, this would place the system outside the far-detuned regime and would generate significant unwanted spin-motion entanglement during the drive. Furthermore, in the presence of any detuning from the COM mode, small (but non-zero) couplings to the remaining modes combine to produce a final interaction graph which is only approximately (but not exactly) the desired all-to-all Hamiltonian.

\begin{figure}[tp]
    \centering\includegraphics{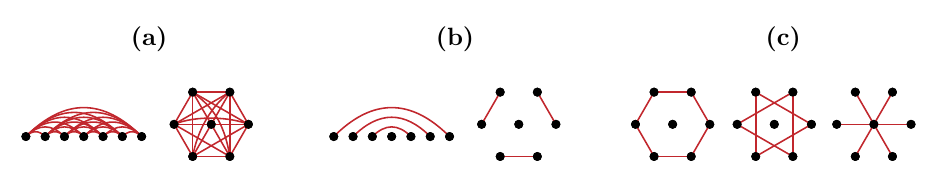}
    \caption{Sample interaction profiles that can be simulated with perfect theoretical fidelity ($\mathcal{I}=0$) for 1D and 2D crystals of $N=7$ ions.
    All connections drawn as red edges have the same strength.
    (a) All-to-all interaction for a 1D chain (left) and a 2D crystal (right).
    (b) Dimer model for a 1D chain (left) and a 7-ion 2D crystal (right).
    The 1D case scales to arbitrary $N$, with the central ion decoupled from the rest for odd $N$.
    The 2D case has perfect fidelity for geometries with a single ion in the center and an even number of ions contained in surrounding circular rings.
    (c) The connections within a 2D triangular-lattice crystal of $N=7$ may be modified using global beams to yield ring, decoupled trimer, and star-like interactions.
    The sum of these three graphs in (c) is equivalent to the all-to-all model from (a).
    }
    \label{fig:4}
\end{figure}

The all-to-all interaction graph can be simulated with perfect fidelity ($\I=0$) if only the COM $\jk$ matrix contributes. This may be accomplished either by setting all weights $c_k$ except the COM to zero, or following equation (\ref{eq:JkProperty}), by setting the COM weight $c_k$ to zero and equal weights $c_k \neq 0$ for all other modes. Experimentally, adding at least $2N+1$ frequency components to the global beams, as in \cite{Shapira2020}, would allow for the decoupling of ion motion from the applied spin interactions and thus enable implementation of these desired mode weights $\{c_k\}$.

The all-to-all interaction may be realized in both 1D ion chains or 2D crystals (figure~\ref{fig:4}(a)), since it requires coupling to only the \com mode. Indeed, \emph{any} ion geometry with a set of transverse motional modes can support an exact all-to-all interaction, independent of geometry or ion number. This follows directly from equation (\ref{eq:Aij}), where it can be seen that a COM mode (with mode vector $\vec{b}_\text{COM}=\{1,1,\ldots 1\}$) is always an eigenvector of the $A_{ij}$ matrix. The existence of this COM mode then permits implementation of an exact all-to-all interaction following the approach outlined above.

\subsubsection{Interacting Dimer Model}
\label{sec:dimer}
Another graph that can be simulated with perfect theoretical fidelity is a collection of non-interacting dimers (figure~\ref{fig:4}(b)). For 1D ion chains, this exactly realizable model scales to arbitrary chain sizes, with an uncoupled ion at the center if $N$ is odd. These dimerized interactions can be generated by selectively driving the spatially symmetric normal modes, or alternatively (following equation (\ref{eq:JkProperty})), by only coupling to the anti-symmetric ones. In addition, this interaction graph for small $N$ can be generated within 2D crystals, as shown in figure~\ref{fig:4}(b), right.

Since any linear combination of accessible interaction graphs is also accessible, combinations of the all-to-all interaction with the interacting dimer model can generate novel connections between spins. For instance, in \cite{shapira2023programmable}, the authors produce a 4-ion ring graph by subtracting two interacting dimers from an all-to-all coupling. Likewise, in \cite{wu2023qubits}, the technique of driving multiple modes is extended to 6- and 8-ion systems to generate nearest-neighbor interactions on a sphere and a hypersphere, respectively. Such linear combinations of exactly solvable models illustrate the flexibility of interactions which are accessible with global beams.

\subsubsection{Modified 2D Lattices}
For two-dimensional ion crystals, driving appropriate sets of transverse vibrational modes enables modification of the native triangular-lattice geometry. For example, the interactions within the 7-ion crystals shown in figure~\ref{fig:4}(c) may be engineered to yield a ring, two decoupled triangular plaquettes, or a ``star"-like central spin model. However, crystal symmetry plays an important role for exact realization of an interaction graph; for general ion numbers, which contain many dislocations within the 2D lattice bulk \cite{kiesenhofer2023controlling}, it is unlikely that any choice of mode weights $\{c_k\}$ will yield a perfectly symmetric modified graph.

As in the one-dimensional case above, the ability to make new graphs by combining sets of other accessible graphs allows for broadened applications. For example, adding the star graph to the ring graph in figure~\ref{fig:4}(c) makes for a system where the central spin is highly frustrated and the ground state is highly entangled. Alternatively, the combination of the 2D dimer graph from figure~\ref{fig:4}(b) and the trimer from figure~\ref{fig:4}(c) is isomorphic to a 3D triangular prism graph, with a tunable ratio of intra-base to inter-base coupling strengths.

%%%%%%%%%%%%%%%%%%%%%%%%%%%%%%%%%%

\subsection{Example Applications: Approximately Realizable Interaction Graphs}

\begin{figure}[tp]
    \centering\includegraphics[width=\textwidth]{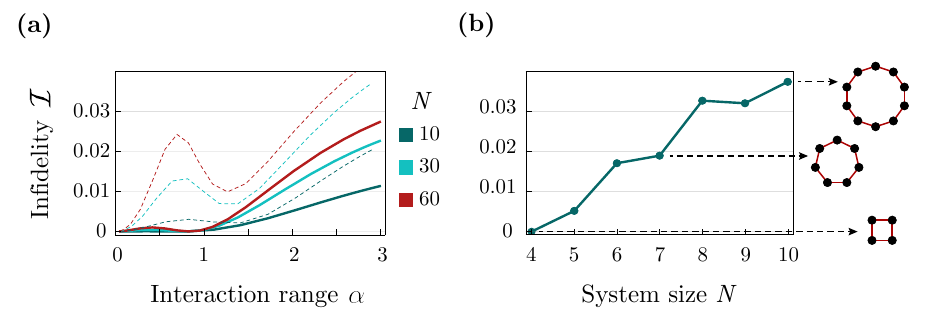}
    \caption{Sample interacting 1D spin models achievable with small infidelity $\mathcal{I}$. (a) Ising-type interactions with power-law decay. Solid lines: infidelity with optimized mode weights $c_k$. Dashed lines: single-beatnote method, same as in figure~\ref{fig:3}, shown here for comparison.
    (b) Minimum infidelity for the ring graph for $N$ vertices (ions), using a linear ion chain.
    For 4 ions, the model is exactly realizable ($\mathcal{I}=0$).
    Vertex relabeling (see Section~\ref{VertexRelabeling}) has been applied to find the minimum infidelity for each $N$.
    }
    \label{fig:5}
\end{figure}

In addition to the exactly realizable models outlined above, we also consider sets of interaction graphs for which the infidelity $\mathcal{I}$ is small but non-zero. For reference, we consider small infidelities to be less than or equal to the typical $1-4\%$ infidelities present in most trapped-ion quantum simulations of long-range Ising models driven with a single bichromatic tone \cite{monroe2021programmable}. We find that by coupling to multiple modes in parallel, numerous interacting spin models may be realized with infidelities at the sub-$1\%$ level.

For example, we first revisit the Ising model with power-law interactions discussed in Section \ref{sec:metrics} and figure \ref{fig:3}. Compared to the typical method of driving with a single bichromatic tone, utilizing multiple modes reduces the infidelity by a factor of approximately 2 to 40, depending on ion number and interaction range $\alpha$ (figure~\ref{fig:5}(a)). It will be shown in Section~\ref{TrappingPotential} that this model's fidelity can be further improved by engineering anharmonic axial confining potentials for the ion crystal. Likewise, nearest-neighbor interactions with periodic boundary conditions (i.e. ring graphs) can also be simulated with low infidelity. As introduced in Section \ref{sec:dimer}, 1D ion chains can generate the nearest-neighbor ring graph exactly for $N=4$ using a linear combination of all-to-all and dimer interactions. While exact solutions are no longer possible for $N>4$ using global beams, figure~\ref{fig:5}(b) demonstrates that the infidelity $\mathcal{I}$ remains low for moderately sized chains.

\begin{figure}[tp]
    \centering\includegraphics[width=\textwidth]{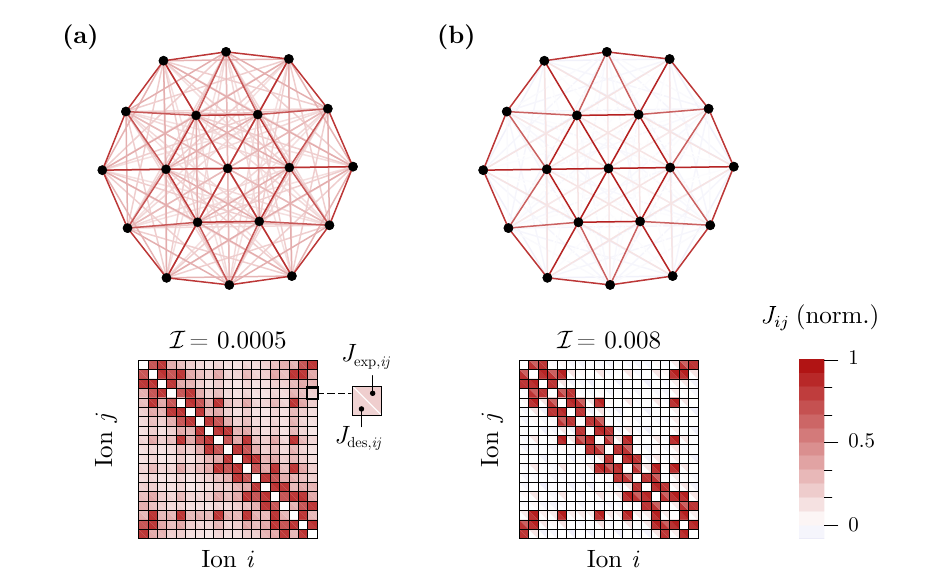}
    \caption{Sample interacting 2D spin models achievable with small infidelity $\mathcal{I}$.
    (a) Ising-type interactions with power-law decay in a 2D crystal of $N=19$ ions, $\{\omega_{x},\omega_y,\omega_z\}=2\pi\times \{5,5,0.1\}$ MHz, and interaction range $\alpha=1.5$. The calculated infidelity is 0.05\%.
    Top: graph of $\jexp$. 
    Bottom: The desired $\jdes$ and the experimentally-achievable $\jexp$ matrix elements shown on the same grid, in the lower and upper triangles (respectively).
    (b) Optimized nearest-neighbor graph and corresponding matrix elements using the same 2D crystal as (a). The calculated infidelity is 0.8\%.
    }
    \label{fig:6}
\end{figure}

Approximately realizable interaction graphs may be generated for 2D ion crystals as well. For a 2D ion crystal, the Ising model with power-law decays may be implemented with $< 1\%$ infidelity over a wide range of ion number and interaction lengths when multiple modes are driven. Figure ~\ref{fig:6}(a) shows the case of $N=19$ and $\alpha=1.5$, which is achievable with 0.05\% infidelity. In addition, 2D crystals support the high-fidelity realization of nearest-neighbor interactions, which are equivalent to a power-law decay model with $\alpha \rightarrow \infty$. Figure~\ref{fig:6}(b) shows an example targeting nearest-neighbor interactions in a triangular lattice of $N=19$ ions, which is achievable with an implementation infidelity of $0.8\%$.

%%%%%%%%%%%%%%%%%%%%%%%%%%%%

\subsection{Vertex relabeling}
\label{VertexRelabeling}

In general, the infidelity $\mathcal{I}$ of a desired interaction graph $\jdes$ may be reduced by relabeling its vertices such that its physical properties are preserved, but it is better adapted to the structure of the ion crystal motional modes.
In graph theory language, an interaction matrix can be thought of as the negative of the Laplacian matrix of a corresponding weighted graph \cite{West2001}, as long as its diagonal entries are chosen to null the sum of each row and column.
Each ion is a vertex, and each spin-spin coupling is an edge with a weight corresponding to the strength of that interaction, including its sign.
Changing the labels of the vertices does not change the physical meaning of that interaction graph. 
However, this relabeled (``isomorphic'') graph has a different Laplacian matrix, and its best experimental approximation $\jexp$ is different in general.
It is therefore advantageous to consider all possible isomorphic graphs with relabeled vertices so that the one with highest fidelity may be selected for experimental implementation.

\begin{figure}[tp]
    \centering\includegraphics[width=\textwidth]{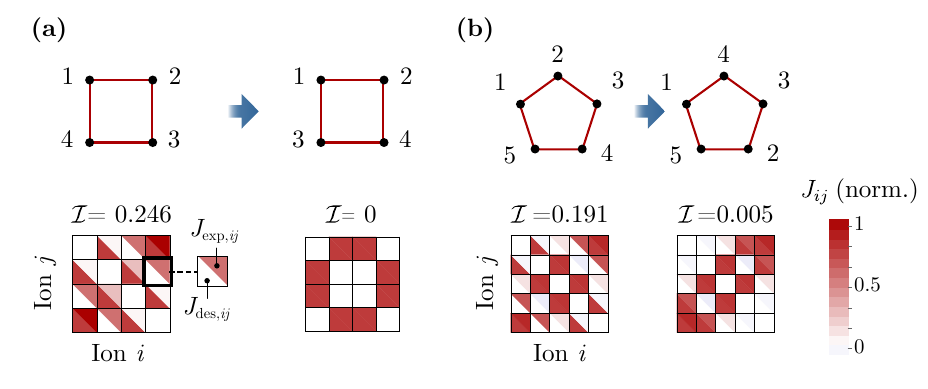}
    \caption{Vertex relabeling can lead to significantly smaller infidelity $\mathcal{I}$.
    (a) For $N=4$ and (b) $N=5$ ions, relabeling vertices leads to a reduction of infidelity from $\sim 20\%$ to $< 1\%$.
    For each $N$, the initial choice of the desired graph and $\jdes$ is shown at the left, and at the right are the optimal ones after vertex relabeling. }
    \label{fig:7}
\end{figure}

An example of such an advantage is shown in figure~\ref{fig:7} for ring graphs with $N=4$ and $N=5$ ions. In each case we begin with an intuitively drawn graph where vertex indices increase monotonically around the ring, corresponding to the simple labeling of ions in a 1D chain from left to right. However, in both cases such connection graphs are only poorly implementable using global beams, with infidelities $\mathcal{I}$ of $\sim 20\%$. Relabeling the vertices to better align with the underlying mode symmetries, as shown in figure~\ref{fig:7}, unlocks the ability to implement these graphs with perfect or near-perfect fidelity.

Formally, vertex relabeling corresponds to the action of a permutation matrix $P$ on the initial desired matrix:
\begin{equation}
P\jdes P^T = \jdes'.
\end{equation}
To find the optimal $\jdes'$, the brute force approach is to apply all $N!$ possible permutations $P$ to the vertex labels and choose the one that leads to the smallest implementation infidelity.
Unfortunately this method scales poorly, with approximately 4 million possible permutations for $N=10$.
While an efficient algorithm to this end is beyond the scope of this paper, we suggest that computational speedups may yet be realized. For instance, the symmetry of the ion motional modes dictates that a high-fidelity solution $\jdes'$ should be symmetric across its anti-diagonal; permutations that strongly violate this condition should be discarded. Applying this straightforward constraint significantly reduces the number of permutations which must be tested before finding an optimal result.

%%%%%%%%%%%%%%%%%%%

\section{Expanded set of interaction profiles using shaped potentials}
\label{TrappingPotential}

In Section~\ref{FromLaserIonInteractionsToSpinSpinInteractions} we showed that accessible interaction graphs are determined by the underlying mode interaction matrices $\jk$, which themselves arise due to the ions' motional modes within the trap potential.
Most commonly, the potential along all three spatial axes is well approximated by a single quadratic (harmonic) term.
In this section we examine what might be gained by allowing the axial confinement to contain anharmonic contributions to the potential, while keeping the transverse directions harmonically confining. We focus on linear chains and investigate two axial anharmonic potentials as examples. In both cases, the extra degrees of freedom gained by shaping the anharmonic terms allow for improved or expanded quantum simulation capabilities using global beams. 

\subsection{Equispaced 1D Ion Chains}

\begin{figure*}[tp]
\centering
\includegraphics[width=\textwidth]{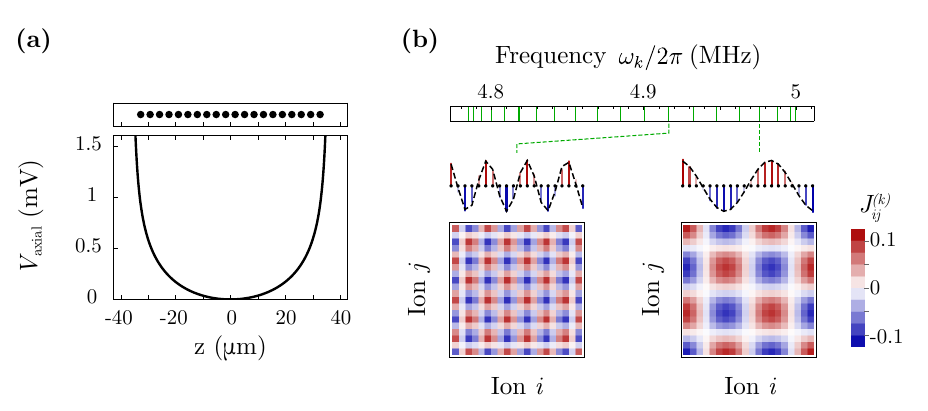}
\caption{(a) Anharmonic axial potential $V_\text{axial}$ with $\tilde{\omega}_{z}=2\pi\times 0.1$ MHz. For $N=20$ ions, this potential leads to a uniform spacing of ions (shown above). (b) The spectrum of transverse normal modes is modified by the anharmonic axial potential. Two mode vectors $\vec{b}_k$ (colored bars) are plotted overlaid with the sinusoidal approximation of equation (\ref{eq:sinmodes}) (dashed lines), showing near-perfect agreement. The corresponding $\jk$ matrices, displayed below, inherit the sinusoidal structure of the mode vectors.
}
\label{fig:8}
\end{figure*}

When a 1D ion chain is confined within a harmonic axial potential, the balance of Coulomb interactions and the trapping fields leads to a clustering of ions near the center of the well. Here, we consider adding higher-order terms to the potential such that the inter-ion spacings are as uniform as possible, as shown in figure \ref{fig:8}(a) \cite{Johanning2016}. Such equispaced ion chains have been investigated previously due to their potential advantages for cooling, computation, and detection \cite{Porras2004,Lin2009,Lin2016,Liu2014,pagano2018cryogenic,Egan2021}.

Beyond these features, equispaced ion chains exhibit a fundamental property which makes them advantageous for improved interaction-graph engineering: their transverse mode vectors $B_{ik}$ are well approximated by the sinusoidal functions
\begin{equation}
\label{eq:sinmodes}
B_{ik} \approx \sqrt{\frac{2-\delta_{k,1}}{N}} \cos \frac{(2i-1)(k-1)\pi}{2N}.
\end{equation}
Figure \ref{fig:8}(b) shows a comparison between the exact normal mode vectors $B_{ik}$ of a 20-ion equispaced chain and the sinusoidal-mode approximation of Eq. \ref{eq:sinmodes}, with discrepancies at the level of $\approx 1\%$. As will be demonstrated below (and proven in \ref{app:dimer} \& \ref{app:transverse}), systems with sinusoidal modes enable new types of interaction graphs to be implemented with perfect theoretical fidelity, while also maintaining the exact realization of all-to-all and dimer interactions introduced in Section \ref{ExactlyRealizableGraphs}. 

The result that equispaced ions lead to sinusoidal modes arises from the structure of the $A_{ij}$ matrix (equation \ref{eq:Aij}) when $|u_i-u_j| \propto |i-j|$.
For equispaced chains, the $d^{\text{th}}$-subdiagonal of the $A_{ij}$ matrix has a constant value inversely proportional to $d^3$, and the diagonal is approximately constant.
This structure is reminiscent of the scenario of a series of harmonic oscillators coupled with springs, providing an intuitive explanation for the sinusoidal form of the eigenvectors in (\ref{eq:sinmodes}).
In the limit $N\rightarrow \infty$, the diagonal is exactly constant and the $A_{ij}$ matrix takes on an infinite Toeplitz form which is known to have sinusoidal eigenvectors \cite{dai2009asymptotics}. (The observation that equispaced ion chains exhibit sinusoidal modes in the $N\rightarrow\infty$ limit was also highlighted in Ref. \cite{Johanning2016}). We note that these results are \emph{not} applicable for the central ions of a harmonically confined chain, which have often been used as a proxy when equal spacing is desired; true anharmonic confinement is required to obtain the mode vector properties discussed above. 

In both the finite and infinite limits, equispaced ion chains can support the same exactly realizable interaction graphs as discussed for harmonic axial confinement in Section \ref{multimode}. Two shared properties between the harmonic and anharmonic potentials lead to this result. First, the highest-frequency transverse mode in both cases is an equal-amplitude COM motion; this guarantees that equispaced chains can admit all-to-all spin-spin interactions with theoretically perfect fidelity. Second, the modes of both harmonically confined and equispaced chains alternate between spatially symmetric and anti-symmetric; equal coupling to all modes with a given parity then produces the same dimer-type interactions as in Sec. \ref{sec:dimer}.

\begin{figure*}[tp]
\centering
\includegraphics[width=\textwidth]{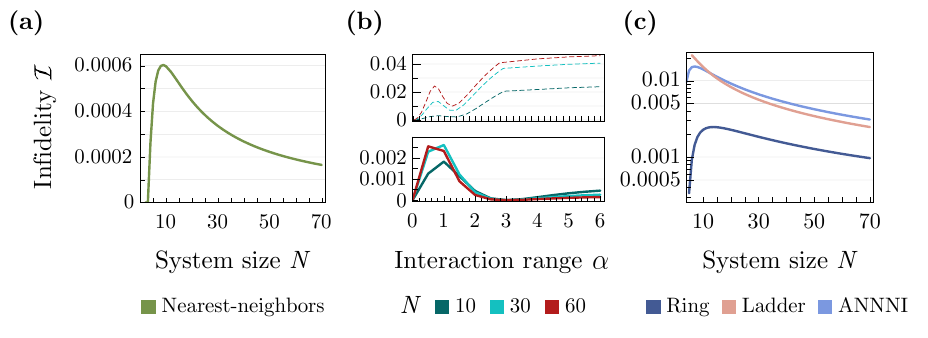}
\caption{Infidelity of interaction models for equispaced 1D chains, using transverse motional modes and a trapping potential as in figure~\ref{fig:8}.
(a) Infidelity of the nearest-neighbor model as a function of system size.
(b) Solid lines: Power-law model infidelity for various system sizes as a function of interaction range $\alpha$.
Compared to the single-beatnote method (dashed lines), equispaced chains yield a smaller infidelity by at least an order of magnitude.
(c) Infidelity of the ring graph, the two-leg rectangular ladder, and the ANNNI model with $J_{i,i+2}=-J_{i,i+1}/2$.
%
%As explained in the main text, the increase of fidelity with system size stems from effectively a single-entry discrepancy of $\jexp$ from $\jdes$, which proportionally affects the fidelity less and less as system sizes increase.
}
\label{fig:9}
\end{figure*}

Beyond these examples, equispaced ion chains open the possibility to realize additional interaction graphs with near-perfect or exactly perfect fidelities in the finite and infinite-ion limits, respectively. For example, figure \ref{fig:9}(a) shows the infidelity for realizing a nearest-neighbor spin-spin interaction within a 1D chain of equispaced ions. Interestingly, the theoretical fidelity of implementing this spin model \emph{improves} for large system sizes. This is a consequence of the underlying mode structure: perfectly-nearest-neighbor interactions can be generated by sinusoidal modes (as proven in \ref{app:transverse}), which are better and better approximated by the ion chain modes in the large-$N$ limit.

Once more, we revisit the Ising model with power-law interactions and compare the performance of an equispaced chain to the standard approach first highlighted in figure \ref{fig:3}. When combining shaped potentials with the multi-mode driving methods of Section \ref{multimode}, we find that the infidelity of power-law Ising interactions may be reduced by one to two orders of magnitude compared to the single bichromatic tone method (figure \ref{fig:9}(b)). In addition, unlike the single-tone approach, we observe that power-law interactions which decay faster than $\sim 1/r^3$ are accessible with low infidelity using anharmonic potentials. Such improved flexibility would allow the study of, for instance, interacting dipolar and van der Waals systems \cite{de2013nonequilibrium,barredo2015coherent,labuhn2016tunable} which were previously inaccessible to trapped-ion simulators with global beams.

While an exhaustive list of spin models realizable with high fidelity using equispaced ions is beyond the scope of this section, we conclude with three additional examples in figure \ref{fig:9}(c). First, we reconsider the ring topology introduced in figure \ref{fig:5}(b). Utilizing the modes of an equispaced ion chain again yields over an order-of-magnitude reduction in the infidelity, with $\mathcal{I} < 0.004$ for any number of ions. Next, we show comparably low infidelities for a spin-ladder geometry which replicates a 2D lattice using a 1D equispaced ion chain. (We note that perfect theoretical fidelity may be achieved following the method of \cite{manovitz2020quantum}, where an uncoupled ``spacer" ion is placed at the center of the chain). Finally, low infidelities may be realized for the Anisotropic Next-Nearest Neighbor Ising (ANNNI) model \cite{fisher1980infinitely}, which is known to exhibit complex phase diagrams and large frustration due to competing nearest- and next-nearest-neighbor interactions. 

%%%%%%%%%%%%%%
%%%%%%%%%%%%%%
%%%%%%%%%%%%%%
\subsection{Double-Well Potentials}

\begin{figure*}[tp]
\centering
\includegraphics[width=\textwidth]{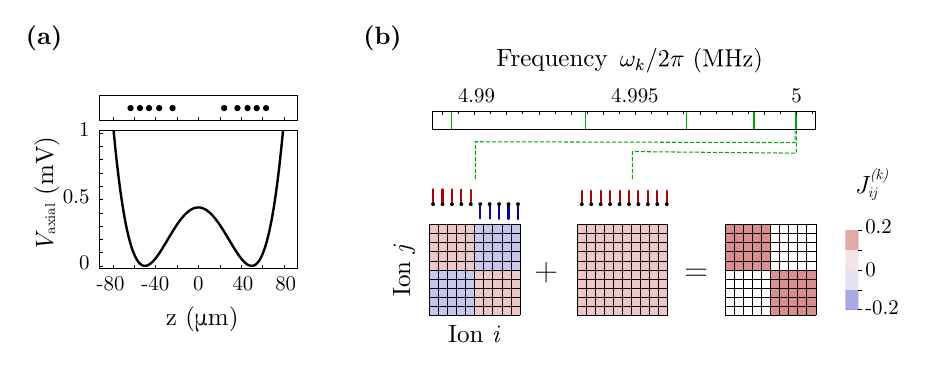}
\caption{(a) Double-well axial potential $V_\text{axial}$ with a confining (positive) quartic term and an anti-confining (negative) quadratic term.
For sufficiently large central barrier, the ion chain is split into two approximately decoupled groups.
(b) At this limit of large separation between the left and right ion groups, the spectrum of transverse normal modes features $N/2$ pairs of modes.
Each near-degenerate pair consists of an even-symmetric mode and its odd-symmetric sibling, which contains opposite signs of ion participation amplitudes in the left and right halves of the chain.
Driving both pairs results in a near-perfect decoupling of interactions between the wells.
}
\label{fig:10}
\end{figure*}

To further illustrate how anharmonic axial potentials may enable new interaction graphs, we consider the case of a double-well potential (figure~\ref{fig:10}(a)). In this configuration, which may be engineered by applying positive voltages to a central set of electrodes, the chain separates into two sets of ions with large Coulomb interactions within each well and small Coulomb interactions across wells.
Here we consider a double well created by combining a confining quartic term with an anti-confining quadratic term, though generic implementations featuring a central ``bulge'' will show the same qualitative features.

In double-well potentials, the transverse modes of the chain lead to a nearly-perfect decoupling of interactions across the central barrier. For $N$ ions, the transverse mode spectrum contains $N/2$ pairs of frequencies, each of which is approximately degenerate (figure \ref{fig:10}(b)), with perfect degeneracy reached in the limit of infinitely separated wells. For the experimentally feasible configuration shown in figure \ref{fig:10}, the frequency separation between mode pairs is calculated to be 40 Hz for the COM mode, and $< 1$~Hz for the zig-zag mode. At each frequency, one of the near-degenerate mode eigenvectors exhibits in-phase motion between ions in separate wells while the other exhibits out-of-phase motion between the wells. Since the modes appear in pairs, any driving laser tone will couple near-equally to both the in-phase and out-of-phase motions. The result is a block-diagonal interaction matrix, where inter-well interactions are canceled and intra-well interactions are dictated by the local mode structure.

For double-well potentials, illuminating the whole chain with the same beam generates the same quantum dynamics for two identical sets of effective spins. Parallel wavefunction evolution (using local addressing) has already found utility in measurements of the second-order R\'enyi entropy to quantify bipartite entanglement in many-body systems \cite{islam2015measuring,linke2018measuring}; here, the state preparation could be performed exclusively using global beams. More generally, if the separately-evolving wavefunctions in both wells are later brought together by reducing or eliminating the central barrier, they may be used to simulate complex systems in materials science and chemistry. For instance, strong local interactions (within each well) followed by relatively weaker interactions between wells takes the form of a Matrix Product State, which has been used in condensed matter physics to describe 1D systems with limited entanglement as well as quasi-2D systems \cite{Vidal2003-od,Schollwock2011-bl}. Furthermore, such setups of weak interactions between strongly coupled 1D systems closely replicate the behavior of chemical nuclear dynamics with multiple interacting nuclear degrees of freedom \cite{saha2021mapping, kumar2022graph}, allowing for extensions to the 1D chemical dynamics simulations already performed with trapped ions \cite{richerme2023quantum}.

%%%%%%%%%%%%%%
%%%%%%%%%%%%%%
%%%%%%%%%%%%%%

\section{Discussion and Outlook}
\label{challenges}
\begin{figure*}[tp]
\centering
\includegraphics[width=\textwidth]{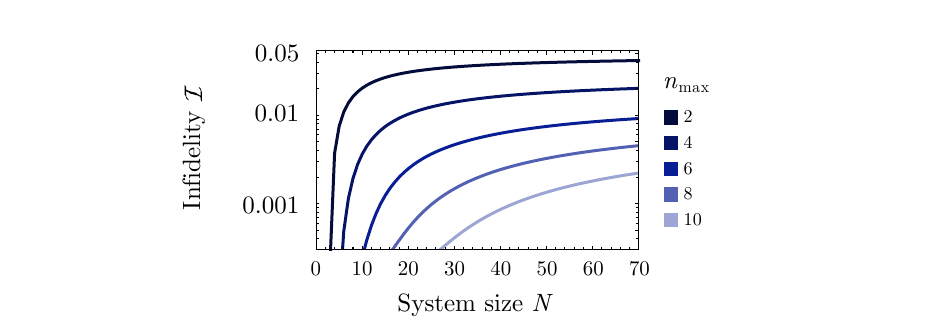}
\caption{Infidelity of realizing the nearest-neighbor model in a 1D chain of $N$ ions, confined in an axial potential with terms up to polynomial order $n_\text{max}$ (equation (\ref{eq:potential})). Shaping the anharmonic confinement with more degrees of freedom leads to lower infidelity. For all ion numbers $N \leq 70$, infidelities of $\mathcal{I} < 0.01$ are achievable using potentials with $n_\text{max} = 6$ (requiring 3 sets of independently controlled dc electrodes).
}
\label{fig:11}
\end{figure*}

In this work, we have described a variety of interaction graphs which are implementable with global beams and have analyzed their maximum possible theoretical performance via the infidelity metric $\mathcal{I}$. This approach abstracts away from common sources of experimental noise, such as trapped-ion heating or photon scattering, which limit the performance of trapped-ion quantum gates \cite{gaebler2016high,ballance2016high}. Thus the total infidelity of implementing the spin-spin interactions described above will depend on both the theoretical minimum infidelity $\mathcal{I}$, as well as the experimental errors specific to each apparatus.

Nevertheless, we highlight two experimental considerations of primary importance for reducing errors during quantum simulations with engineered interaction graphs. First, as described in Section \ref{multimode}, successful interaction engineering relies upon an appropriately weighted sum of $\jk$ matrices, each with weight $c_k$. Following the methods in Refs. \cite{shapira2018robust,Shapira2020}, arbitrary weights may be generated via the application of $2N+1$ bichromatic beatnotes $\mu_m$, each with their own Rabi frequency $\Omega_m$. However, this suggests that motional-mode frequency drifts (due to drifts in the rf voltage and/or frequency, for instance), or laser intensity fluctuations (due to power fluctuations or pointing instability), will have outsized effects in limiting the experimental fidelity. Beyond the straightforward mitigation strategies of rf \cite{johnson2016active} and laser stabilization, we note that even more bichromatic beatnotes may be added to reduce the sensitivity of mode weights to frequency and intensity fluctuations. In Ref. \cite{shapira2023robust}, for instance, the authors demonstrate that additional beatnotes may be used to reduce the infidelity of a standard M{\o}lmer-S{\o}rensen gate by nearly an order of magnitude in the presence of $\sim 5\%$ amplitude and mode frequency errors.

Second, as described in Section \ref{TrappingPotential}, utilizing strings of equispaced ions further improves or expands upon the quantum simulation possibilities achievable with global beams compared to the harmonic case. However, following the analysis presented in \cite{Johanning2016}, perfectly equispaced ion chains are idealized since they require control over an infinite number of anharmonic potential terms. Motivated by prior work such as \cite{Lin2009}, which achieve nearly equispaced chains by controlling only the second and fourth order terms in the axial potential, we quantify the expected experimental errors arising from non-uniform ion spacings. Figure \ref{fig:11} considers the infidelity $\mathcal{I}$ of realizing nearest-neighbor interactions in a 1D chain of $N$ ions, when only terms up to polynomial order $n_\text{max}$ are used to shape the axial potential (equation (\ref{eq:potential})). We find that only the first few potential orders are required to keep the infidelity at or below the $1\%$ level, even for large system sizes, and remark that a symmetric potential of order $n_\text{max}$ may be implemented within an ion trap using $n_\text{max}/2$ sets of symmetric dc electrodes.

We have presented a suite of techniques for expanding the reach of quantum simulations using exclusively global beams. We have shown that by driving all available vibrational modes with appropriate weights, previously inaccessible spin-spin coupling graphs become implementable in trapped-ion simulators with perfect or near-perfect fidelity. We developed a simple but rigorous test to determine whether a desired interaction profile may be perfectly mapped to a trapped-ion system given its set of vibrational normal modes. Additionally, we showed that further high-fidelity classes of spin-spin interactions become achievable by considering shaped anharmonic axial potentials. Taken together, these tools make a wide range of new problems in materials science and chemistry accessible to ion-trap quantum simulators, while avoiding the experimental overhead and complexity associated with locally addressed entangling interactions. 

\section*{Acknowledgements}
The authors thank Yotam Shapira for initial discussions. This material is based upon work supported by the National Science Foundation under Grants No. OMA-1936353 and CHE-2311165. The IU Quantum Science and Engineering Center is supported by the Office of the IU Bloomington Vice Provost for Research through its Emerging Areas of Research program.

%%%%%%%%%%%%%%%%%%%%%%%%%
%%%%%%%%%%%%%%%%%%%%%%%%%%%%

\section*{References}
\bibliographystyle{prsty}
\bibliography{refs}

%%%%%%%%%%%%%%
%%%%%%%%%%%%%%
%%%%%%%%%%%%%%

\appendix

%%%%%%%%%%%%%%
%%%%%%%%%%%%%%
%%%%%%%%%%%%%%

\section{Proof of equation (\ref{eq:JkProperty})}
\label{app:proofofjk}
The $N\times N$ mode vector matrix $B$ is the modal matrix of $A$, which is real and symmetric. As such, $B$ is orthonormal. One of its properties is then
\begin{equation}\label{eq:BBT}
BB^T = B^TB = I.
\end{equation}
or
\begin{equation}
(BB^T)_{ij} = \delta_{ij}.
\end{equation}
We re-write this as,
\[
(BB^T)_{ij} = \sum_k B_{ik}(B^T)_{kj} = \sum_k B_{ik} B_{jk}
\]
and using (\ref{eq:jkDefinition}) this becomes:
\[
(BB^T)_{ij} = \sum_k J_{ik}^{(k)} \Rightarrow \delta_{ij} = \sum_k J_{ik}^{(k)}
\]
or in matrix form
\[
\sum_k \jk = \mathbf{1}.
\]

%%%%%%%%%%%%%%%%%%%%%%%%%%%%%%%%%

\section{Proof of equation (\ref{eq:Centerpiece}) for harmonic confinement}
\label{app:AccessibilityProof}

In this section we provide the proof for equation (\ref{eq:Centerpiece}). Let us assume that a desired interaction matrix $\jdes$ can be decomposed as a linear combination of $\jk$'s:
\begin{align}
\label{eq:qwer}
\jdes &= \sum_k c_k J^{(k)},\ c_k\in \mathbb{R}\\
\Leftrightarrow J_{\text{des},ij} &= \sum_k c_k B_{ik} B_{jk}\\
\Leftrightarrow J_{\text{des},ij}&= \sum_k c_k B_{ik} (B^T)_{kj}\\
\Leftrightarrow J_{\text{des},ij}&= \sum_k \sum_l B_{ik} C_{kl} (B^T)_{lj}
\end{align}
where $C$ is a diagonal matrix with the weights $c_k$ in its diagonal:
\begin{equation}
C\equiv \diag(\{c_k\})
\end{equation}
In matrix form, (\ref{eq:qwer}) reads
\begin{equation}
\jdes = BCB^T \Leftrightarrow B^T\jdes B = C
\end{equation}
We note that with the convention that $\jdes$ is provided as input in the graph Laplacian form, i.e. with its diagonal chosen to zero the sum of each row and column, the set of $\{c_k\}$ in the relations above is unique, and the \com mode's weight is always zero: $c_\text{COM}=0$.
However, the diagonal of an interaction matrix bears no physical significance.
This provides the experimental flexibility, after a set of mode weights $\{c_k\}$ has been found, to add a constant to all of them, without changing the physical interaction matrix, such that for example the \com mode's role is diminished, since it is especially prone to ion heating.

%%%%%%%%%%%%%%%%
%%%%%%%%%%%%%%%%
\section{Accessibility of the dimer model}
\label{app:dimer}
%%%%%%%%%%%%%

We start by showing that the dimer model is accessible for sinusoidal transverse modes given by equation (\ref{eq:sinmodes}):
\begin{equation}\label{eq:sinmodes2}
B_{jk} = \sqrt{\frac{2-\delta_{k,1}}{N}} \cos \frac{(2j-1)(k-1)\pi}{2N}.
\end{equation}
The mode interaction matrices then take the form
\begin{equation}
J^{(k),\sin}_{ij} = \frac{2-\delta_{k,1}}{N} \cos \frac{(2i-1)(k-1)\pi}{2N} \cos \frac{(2j-1)(k-1)\pi}{2N}.
\end{equation}

We will show that coupling equally to every other mode results in the pair-wise interaction model shown in figure~\ref{fig:4}(b).
To this end, we will calculate the sum of the mode interaction matrices, assuming $N$ is even for simplicity
\begin{equation}
\mathbb{J} \equiv J^{(2)} + J^{(4)} + J^{(6)} +\ldots
\end{equation}
which is
\begin{align}
\mathbb{J} &= \frac{2}{N}\sum_{k=2,4,6,\ldots}^N \cos\frac{(2i-1)(k-1)\pi}{2N} \cos\frac{(2j-1)(k-1)\pi}{2N} \\
&=\frac{2}{N}\sum_{k=1,3,5,\ldots}^{N-1} \cos\frac{(2i-1)k\pi}{2N} \cos\frac{(2j-1)k \pi}{2N}
\end{align}
%where by $N_\text{even}$ we denote the even number between $\{N-1,N\}$ and similarly for $N_\text{odd}$.
Using the  product-of-cosines identity,
\begin{equation}\label{eq:whatev}
\mathbb{J} = \frac{1}{N}\sum_{k\text{ odd}}^{N-1} \cos \frac{k\pi}{N}(i+j-1) + \cos \frac{k\pi}{N}(i-j)
\end{equation}
We will now use the identity
\begin{equation}\label{eq:lagrrr}
\sum_{\kappa=1}^n \cos (2\kappa-1)a = \frac{\sin 2 n a}{2\sin a},\quad a\neq l\pi,l\in\mathbb{Z},
\end{equation}
re-written as
\begin{equation}
\sum_{k=1,\text{odd}}^{N-1} \cos k a = \frac{\sin N a}{2\sin a},\quad a\neq l\pi,l\in\mathbb{Z}.
\end{equation}
for each of the two terms inside the summation in (\ref{eq:whatev}),
with $a\rightarrow\pi(i+j-1)/N$ for the first term and $a\rightarrow\pi(i-j)/N$ for the second term.
We note that the only values of $i,j\in[1,N]$ that these $a$'s can be equal to $l\pi,\ l\in\mathbb{Z}$, are for $l=1$ and $l=0$ respectively.
So, the values of $i,j$ for which the identity (\ref{eq:lagrrr}) does not apply are $i+j-1=N\Rightarrow j=N-i+1$ and $i-j=0\Rightarrow j=i$.
For all $i,j$ except for these values, we have
\begin{equation}\label{eq:zero}
\mathbb{J}_{ij} = \frac{1}{N} \left.\frac{\sin \pi (i+j-1)}{2\sin \frac{\pi (i+j-1)}{N}}\right|_{i+j-1\neq l\pi}  +  \frac{1}{N} \left.\frac{\sin \pi (i-j)}{2\sin \frac{\pi (i-j)}{N}}\right|_{i-j\neq l\pi} = 0
\end{equation}
For the mentioned values of $i,j$ where identity (\ref{eq:lagrrr}) does not hold, we have

\begin{itemize}
\item \underline{$j=N-i+1$ (i.e., the anti-diagonal of $\mathbb{J}$)}
\end{itemize}
\begin{align}
\mathbb{J}_{ij}|_{j=N-i+1} &= \frac{1}{N} \sum_{k\text{ odd}}^{N-1} \left[\underbrace{\cos k\frac{\pi}{N}0}_{=1} + \cos k\frac{\pi}{N}(2i-N-1)\right]\\
&= \frac{1}{N} \left[ \frac{N}{2}+ \frac{\sin [\pi(2i-N-1)]}{2 \sin \pi/N(2i-N-1)} \right]
\end{align}
For even $N$, there are no values of $i$ that zero the sine in the denominator, and we can replace it with zero since its numerator is an integer multiple of $\pi\ \forall i,N$:
\begin{equation}\label{eq:AntiDiag}
\mathbb{J}_{ij}|_{j=N-i+1} = \frac{1}{2}.
\end{equation}

\begin{itemize}
\item \underline{$j=i$ (i.e., the diagonal of $\mathbb{J}$)}
\end{itemize}
\begin{align}
\mathbb{J}_{ij}|_{j=i} &= \frac{1}{N} \sum_{k\text{ odd}}^{N-1} \left[\cos \frac{k\pi}{N}(2i-1) + \underbrace{\cos\frac{k\pi}{N}0}_{=1}\right] \\
&= \frac{1}{N} \left[ \frac{\sin \pi(2i-1)}{2\sin \pi/N(2i-1)} + \frac{N}{2}\right]
\end{align}
For even $N$, there is no value of $i\in[1,N]$ that zeroes the sine in the denominator.
Therefore, similar as before, that term is always zero, and we can write
\begin{equation}\label{eq:Diag}
\mathbb{J}_{ij}|_{j=i} = \frac{1}{2}
\end{equation}
In combination, (\ref{eq:zero}), (\ref{eq:AntiDiag}), and (\ref{eq:Diag}) show that $\mathbb{J}$ is a sparse matrix with values of 1/2 in its diagonoal and anti-diagonal, which is the matrix representation of the interacting dimers model.

The derivation above can be generalized to non-sinusoidal modes of harmonically confined chains, where each mode vector is now a linear combination of the sinusoidal mode vectors.

%%%%%%%%%%%%%%%%%%%%%

\section{Accessibility of the nearest-neighbor model for sinusoidal transverse modes}
\label{app:transverse}

The transverse modes for equispaced chains of size $N$ are well approximated by
\begin{equation}\label{eq:TransvCosModes}
B_{jk} = \sqrt{\frac{2-\delta_{k,1}}{N}} \cos \frac{(2j-1)(k-1)\pi}{2N}.
\end{equation}
Here we show that the mode weights
\begin{equation}
c_k = 2 \cos\frac{(k-1)\pi}{N}\quad,k=1,\ldots,N
\end{equation}
result to a nearest-neighbor model as long as the mode vectors are as in (\ref{eq:TransvCosModes}).
The mode interaction matrices now take the form
\begin{equation}
J^{(k),\sin}_{ij} = \frac{2-\delta_{k,1}}{N} \cos \frac{(2i-1)(k-1)\pi}{2N} \cos \frac{(2j-1)(k-1)\pi}{2N}
\end{equation}
and the proposed mode weights lead to the interaction matrix
\begin{IEEEeqnarray}{rCl}
\mathbb{J} &=& \sum_{k=1}^N c_k J^{(k),\sin}_{ij}\\
&=& \sum_{k=1}^N  \frac{4-2\delta_{k,1}}{N} \cos\frac{(k-1)\pi}{N} \cos \frac{(2i-1)(k-1)\pi}{2N} \cos \frac{(2j-1)(k-1)\pi}{2N}\\
&=& \sum_{k=0}^{N-1}  \frac{4-2\delta_{k,0}}{N} \cos\frac{k\pi}{N} \cos \frac{(2i-1)k\pi}{2N} \cos \frac{(2j-1)k\pi}{2N}\\
&=& \frac{2}{N} + \frac{4}{N} \sum_{k=1}^{N-1}\cos\frac{2k\pi}{2N} \cos \frac{(2i-1)k\pi}{2N} \cos \frac{(2j-1)k\pi}{2N}\\
&=& \frac{2}{N} + \frac{4}{N} \sum_{k=1}^{N-1} \cos \frac{(2i-3)k\pi}{2N} \cos \frac{(2j-1)k\pi}{2N}  +  \cos \frac{(2i+1)k\pi}{2N} \cos \frac{(2j-1)k\pi}{2N}.
\end{IEEEeqnarray}
Using the cosine-cosine product trigonometric identity twice:
\begin{IEEEeqnarray}{rCl}
\mathbb{J}_{ij} &=& \frac{2}{N} + \frac{1}{N}\sum_{k=1}^{N-1} \left[\cos\frac{k(i+j)\pi}{N} + \cos\frac{k(i-j+1)\pi}{N} \right. \\
&&\left. + \cos\frac{k(1-i+j)\pi}{N} + \cos\frac{-k(i+j-2)\pi}{N}\right]\label{eq:whatv}
\end{IEEEeqnarray}
Now we use Lagrange's trigonometric identity for each of the 4 sums in (\ref{eq:whatv}).
We will explicitly replace the first term, and use ``$\ldots$'' for the other three for now.
\begin{IEEEeqnarray}{rCll}
\mathbb{J}_{ij} &=&& \frac{2}{N} + 
\frac{1}{N} \left[-\frac{1}{2} - \cos(i+j)\pi + \frac{\sin [(2N+1)\frac{(i+j)\pi}{2N}]}{2\sin[\frac{(i+j)\pi}{2N}]}  +\ldots   \right] \\
&=&& \frac{2}{N} + \frac{1}{N} \left[-\frac{1}{2} - \cos(i+j)\pi +  \frac{\sin [(1+1/(2N))(i+j)\pi]}{2\sin[(i+j)\pi/(2N)]} +\ldots \right]\label{eq:qwert}\\
&=&& \frac{2}{N} + \frac{1}{N} \left[-\frac{1}{2} - \cos(i+j)\pi + \frac{1}{2}(1 - \delta_{i+j,2N})\cos(i+j)\pi + \frac{1}{2}\delta_{i+j,2N}(2N+1)  +\ldots \right]\nonumber\\
    &=&&\frac{2}{N} + \frac{1}{N} \left[-\frac{1}{2}  - \frac{1}{2}\cos(i+j)\pi  +  \frac{1}{2}\delta_{i+j,2N}(2N+1-\cos(i+j)\pi) + \ldots\right]
\end{IEEEeqnarray}
The first cosine term, plus the rest three homologous terms at the ``\ldots'' part of the sum add to zero for all integers $i,j$.
Also, the terms involving the Kronecker delta are the values of $i,j$ for which the denominator with the sine at (\ref{eq:qwert}) equals zero.
After some algebra, we can write
\begin{equation}
\mathbb{J}_{ij} = \frac{2}{N} + \frac{1}{2N} \left[  -4 + 2N(\delta_{i+j,2N}+ \delta_{i-j+1,0} + \delta_{-i+j+1,0} + \delta_{i+j-2,0} ) \right] 
\end{equation}
or equivalently
\begin{equation}
\mathbb{J}_{ij} = \delta_{i+j,2N} + \delta_{i+j,2} + \delta_{i-j,1} + \delta_{i-j,-1}
\end{equation}
which is the nearest-neighbor matrix, with 2 inconsequential nonzero diagonal entries: $\mathbb{J}_{1,1}$ and $\mathbb{J}_{NN}$.

\end{document}